\RequirePackage{fix-cm}
\documentclass[twocolumn,epjc3]{svjour3}  
\smartqed  
\RequirePackage{graphicx}
\usepackage{caption}
\usepackage{amsmath}
\usepackage{subfig}
\usepackage{braket}
\usepackage{lipsum}
\journalname{}
\begin{document}

\title{Circular polarization measurement for individual gamma rays in capture reactions with intense pulsed neutrons
}


\author{Shunsuke Endo\thanksref{e1,addr1,addr2}
 \and
       Ryota Abe\thanksref{addr2} 
           \and
       Hiroyuki Fujioka\thanksref{addr3} 
       \and
       Takashi Ino\thanksref{addr4}
       \and
        Osamu Iwamoto\thanksref{addr1}
         \and
        Nobuyuki Iwamoto\thanksref{addr1}     
        \and
        Shiori Kawamura\thanksref{addr2}
         \and
        Atsushi Kimura\thanksref{addr1}
        \and
        Masaaki Kitaguchi\thanksref{addr2}
        \and
        Ryujyu Kobayashi\thanksref{addr5}
        \and
        Shoji Nakamura\thanksref{addr1}
        \and
        Takayuki Oku\thanksref{addr1}
         \and
        Takuya Okudaira\thanksref{addr2,addr1}
         \and
        Mao Okuizumi\thanksref{addr2}
        \and 
        Mohamed Omer\thanksref{addr1}
        \and
         Gerard Rovira\thanksref{addr1}
            \and
        Tatsushi Shima\thanksref{addr6}
           \and
       Hirohiko M. Shimizu\thanksref{addr2}
          \and
        Toshiyuki Shizuma\thanksref{addr7}
           \and
       Yoshitaka Taira\thanksref{addr8}
          \and
        Shusuke Takada\thanksref{addr9}
           \and
       Shingo Takahashi\thanksref{addr4}
        \and
        Hiromoto Yoshikawa\thanksref{addr6}
                \and
        Tamaki Yoshioka\thanksref{addr10}
                \and
        Heishun Zen\thanksref{addr11}
}

\thankstext{e1}{e-mail: endo.shunsuke@jaea.go.jp}


\institute{Japan Atomic Energy Agency, 2-4 Shirakata, Tokai 319-1195, Japan \label{addr1}
           \and
            Nagoya University, Furocho, Chikusa, Nagoya 464-8062, Japan \label{addr2}
           \and
            Tokyo Institute of Technology, 2-12 Ookayama, Meguro-ku, Tokyo 152-8511, Japan \label{addr3}
            \and
            High Energy Accelerator Research Organization (KEK), 1-1 Oho, Tsukuba, Ibaraki 305-0801, Japan \label{addr4}
            \and
            Ibaraki University, 2-1-1 Bunkyo, Mito, Ibaraki 310-8512, Japan \label{addr5}
            \and
            Osaka University, 10-1 Mihogaoka, Ibaraki, Osaka 567-0047, Japan \label{addr6}
            \and
            National Institute for Quantum Science and Technology, 8-1-7 Umemidai, Kizugawa-shi, Kyoto 619-0215, Japan \label{addr7}
            \and
            Institute for Molecular Science, Myodaiji, Okazaki, Aichi 444-8585, Japan \label{addr8}
            \and
            Tohoku University, 2-1-1 Katahira, Aoba-ku, Sendai, Miyagi 980-8576, Japan \label{addr9}
            \and
            Kyushu University, 744 Motooka, Nishi-ku, Fukuoka, Fukuoka 819-0395, Japan \label{addr10}
            \and
            Institute of Advanced Energy, Kyoto University, Gokasho, Uji, Kyoto 611-0011, Japan \label{addr11}
}

\date{Received: date / Accepted: date}

\maketitle

\begin{abstract}
Measurements of circular polarization of $\gamma$-ray emitted from neutron capture reactions provide valuable information for nuclear physics studies. The spin and parity of excited states can be determined by measuring the circular polarization from polarized neutron capture reactions. Furthermore, the $\gamma$-ray circular polarization in a neutron capture resonance is crucial for studying the enhancement effect of parity nonconservation in compound nuclei. The $\gamma$-ray circular polarization can be measured using a polarimeter based on magnetic Compton scattering. A polarimeter was constructed, and its performance indicators were evaluated using a circularly polarized $\gamma$-ray beam. Furthermore, as a demonstration, the $\gamma$-ray circular polarization was measured in $^{32}$S($\vec{\textrm{n}}$,$\gamma$)$^{33}$S reactions with polarized neutrons. 

\end{abstract}

\section{Introduction}

Measuring the circular polarization of $\gamma$ rays emitted from nuclear reactions is used to determine spin correlations. In $\gamma$-ray transitions from excited states formed by neutron capture reactions, the circular polarization of each $\gamma$ ray depends on the spin of the excited states and the mixing of the multipolarity in the transitions~\cite{Biedenharn}. Therefore, measuring circular polarization is valuable for studying nuclear structures.

Furthermore, $\gamma$-ray circular polarization in a neutron capture resonance is crucial for studying the enhancement effect of the parity nonconservation (parity-violation) in compound nuclei. A significant enhancement of the parity-violation in the compound nuclear reactions has been observed in longitudinally polarized neutron capture reactions~\cite{Mitchell}. This enhancement results from mixing the s- and p-wave amplitudes, predicted by the s-p mixing model~\cite{Flambaum}. The s-p mixing model also gives a formula for the neutron capture cross-sections considering the angular correlations between neutron momentum, neutron spin, $\gamma$-ray momentum, and $\gamma$-ray spin depending on neutron energy. The angular correlation terms have been measured for several nuclei, such as $^{139}$La, $^{117}$Sn, and $^{131}$Xe ~\cite{Okudaira2018,Yamamoto,Okudaira2021,Koga,Endo2022,Okudaira2023}, to verify the s-p mixing model. These correlation terms result from the interference between the s- and p-wave amplitudes predicted by the s-p mixing model. However, they do not include the parity-violating effect caused by mixing those amplitudes. The circular polarization-dependent cross-section is a correlation term equivalent to the parity-violating cross-section caused by the mixing s- and p-wave amplitudes. If parity violation enhancement is based solely on the s-p mixing model, the magnitude of the parity violation measured by circular polarization should be interpreted in a unified manner with the measurement results using polarized neutrons~\cite{Mitchell}. Therefore, circular polarization measurements are significant for validating the s-p mixing model. None-theless, the circular polarization of $\gamma$ rays from resonances in the high-energy region of approximately 1~eV or more, called epithermal neutrons, has not been measured.

The $\gamma$-ray circular polarization can be determined by measuring the transmission through a magnetized material, called a $\gamma$-ray polarimeter. Several circular polarization measurements in neutron capture reactions were performed from the 1950s to the 1970s~\cite{TRUMPY,Vervier,KOPECKY,Djadali,VENNINK}. The $\gamma$-ray polarimeter comprised an iron magnetized by a wrapped coil. These measurements were performed using low-energy neutrons ($<25$~meV) from a nuclear reactor to determine the spin of a compound state in thermal neutron energy. However, it is challenging to determine the spin uniquely because the effects of many resonances are mixed in the thermal-neutron energy. In the 2000s, a polarimeter was developed to measure the parity-violation in polarized thermal neutrons and deuteron reactions~\cite{MikePol}. The polarimeter comprised permendur material, which has the strongest saturation magnetization among common magnetic materials. The uses of $\gamma$-ray polarimeters were limited to $\gamma$-rays emitted from thermal neutron capture reactions because no facilities provided highly intense epithermal neutrons.

Recently, highly intense epithermal neutrons have become available because of the development of highly intense pulsed neutron sources. Hence, introducing a polarimeter to pulsed neutron sources makes it possible to measure the circular polarization of $\gamma$ rays emitted from resonance states. Therefore, it is possible to perform experiments to determine the spin of resonance states and verify the s-p mixing model. Consequently, we developed a $\gamma$-ray polarimeter that can be used at the Materials and Life Science Experimental Facility (MLF) in the Japan Proton Accelerator Research Complex (J-PARC). The MLF in J-PARC has a pulsed neutron source with one of the highest neutron intensities globally. The Accurate Neutron-Nucleus Reaction Measurement Instrument (ANNRI) is a beamline in the MLF. High-intensity neutrons ranging from a few meV to several keV are available at ANNRI, where (n,$\gamma$) reaction measurements are performed using the installed high-purity germanium (HPGe) detectors. Thus, ANNRI is an ideal instrument for conducting $\gamma$-ray circular polarization measurements that require large statistics.

The polarimeter has two performance indicators: the analyzing power, known as capability in the polarization measurement, and the magnetic hysteresis. The magnetization of the polarimeter must be saturated for circular polarization measurements. In the present study, the $\gamma$-ray transmission was measured as a function of the current applied to the polarimeter using circularly polarized $\gamma$ rays at the Ultraviolet Synchrotron Orbital Radiation  (UVSOR) facility~\cite{Katoh}. The experimental results on the two performance indicators will be presented.


As a demonstration of the circular polarization measurement of the neutron capture reaction in a pulsed neutron source, the circular polarization of $\gamma$ rays was measured in $^{32}$S($\vec{\textrm{n}}$,$\gamma$)$^{33}$S reactions at ANNRI. The circular polarization of several individual $\gamma$ rays was derived and compared with the theoretical calculation. In this paper, we discuss the circular polarization of $\gamma$-rays produced in the polarized neutron capture of $^{32}$S.

\section{Gamma-ray Polarimeter}
The Compton scattering cross-section including magnetic effect is written as~\cite{Chesler,Tolhoek}:
\begin{equation}
\label{eq:magcomp}
    \sigma_\textrm{Comp}=2\pi r_0^2(\sigma_0+P_\textrm{e}P_\gamma\sigma_c\cos\theta),
\end{equation}
where $r_0(=e^2/mc^2)$ is the electron radius, $\sigma_0$ is the Klein-Nishina Compton scattering cross-section, $P_\textrm{e}$ and $P_\gamma$ are the degrees of the electron polarization and the $\gamma$-ray circular polarization, respectively, and $\theta$ is the angle between the electron spin direction and the $\gamma$-ray momentum direction. $\sigma_c$ is expressed as
\begin{equation}
\sigma_c=\frac{1+4\alpha+5\alpha^2}{\alpha(1+2\alpha)^2}-\frac{(1+\alpha)\ln(1+2\alpha)}{2\alpha^2},
\end{equation}
where $\alpha$ is initial photon energy in units of the electron mass. The second term in Eq.~(\ref{eq:magcomp}), representing the magnetic effect, depends on the $\gamma$-ray circular polarization. Thus, circular polarization can be determined by measuring the $\gamma$-ray transmission of a polarimeter.

\begin{figure}[htbp]
	\centering
		\includegraphics[clip,width=8cm]{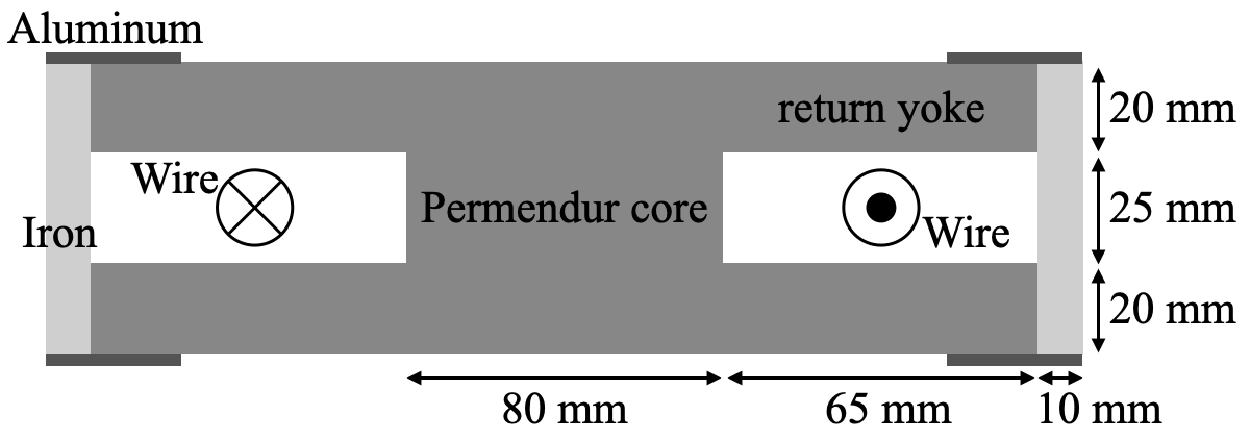}
	  \caption{A cross-sectional view of the polarimeter. The core comprised permendur, and the return yoke comprised permendur and iron.}\label{polarimeter}
\end{figure}

Several experiments~\cite{TRUMPY,KOPECKY,VENNINK,MikePol} have used the magnetic Compton scattering to measure $\gamma$-ray circular polarization. A transmission polarimeter was constructed in a manner similar to that used in these studies. However, the size was limited to installation in ANNRI. To improve the analyzing power, the core of the polarimeter comprised a permendur material that provided a high saturation magnetization corresponding to high-electron polarization. Furthermore, the polarimeter included a return yoke to reduce the effect of the magnetic field leakage on the HPGe detector and achieve a high magnetic field. The polarimeter was wound with 500~turns of conductor and operated at a current of 3.98~A. Figure~\ref{polarimeter} shows a sectional view of the polarimeter.
The magnetic field was simulated using computer-aided engineering (CAE) software, Femtet, produced by Murata Software Co., Ltd.~\cite{Femtet}. The average magnetization of the permendur core was estimated to be $1.34$~T at $3.98$~A.

\begin{figure}[htbp]
	\centering
		\includegraphics[clip,width=8cm]{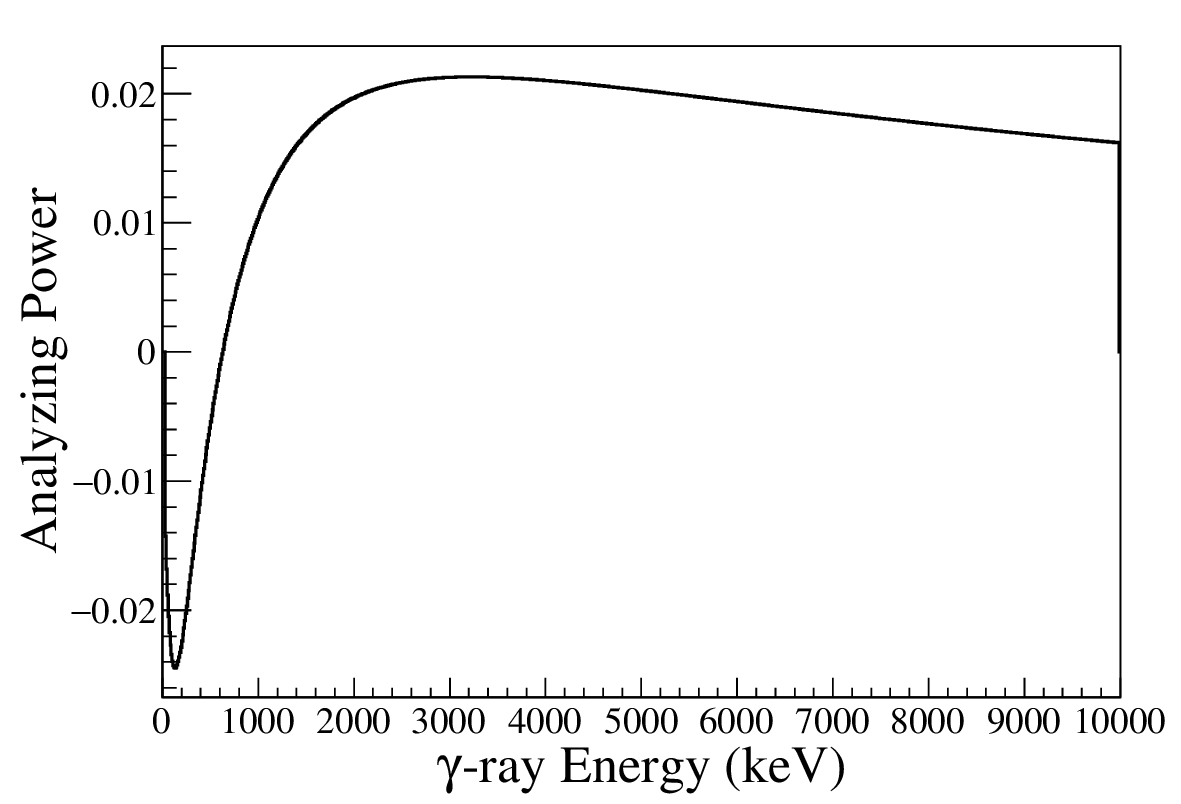}
	  \caption{Calculated $\gamma$-ray energy dependence of the analyzing power.}\label{Edep_Pa}
\end{figure}

The analyzing power for the $\gamma$-ray circular polarization is calculated as
\begin{equation}
\label{eq:Padef1}
    P_a=\frac{T_\textrm{r}-T_\textrm{l}}{T_\textrm{r}+T_\textrm{l}},
\end{equation}
where $T_\textrm{r}$ and $T_\textrm{l}$ are the transmissions for right- and left-handed circularly polarized $\gamma$ rays.
The transmission is calculated as $T_\textrm{r}=\exp(-n_0L\sigma_\textrm{r})$, where $n_0$ is the number density of the electrons, $L$ is the core length, and
\begin{eqnarray}
\sigma_\textrm{r}=(\sigma_\textrm{PE}+2\pi r_0^2(\sigma_0+P_\textrm{e}\sigma_\textrm{c})+\sigma_\textrm{Co}+\sigma_\textrm{PP})N_A/A, \nonumber\\
\sigma_\textrm{l}=(\sigma_\textrm{PE}+2\pi r_0^2(\sigma_0- P_\textrm{e}\sigma_\textrm{c})+\sigma_\textrm{Co}+\sigma_\textrm{PP})N_A/A.
\end{eqnarray}
Here, $\sigma_\textrm{PE}$, $\sigma_\textrm{Co}$, and $\sigma_\textrm{PP}$ are the cross-sections of the photoelectron effect, coherent scattering, and pair production, respectively. $N_A$ is the Avogadro constant, and $A$ is the atomic weight. The magnetization of the polarimeter, $M$, is related to the electron polarization ratio ~\cite{Aulenbacher}:
\begin{equation}
    P_\textrm{e}=M\frac{(g'-1)2}{(g-1)g'}\frac{1}{N_\textrm{e}\mu_\textrm{B}},
\end{equation}
where $\mu_\textrm{B}$ is the Bohr magneton, $g$ is the electron g-factor ($g=2.002$), $g'$ is the g-factor for the permendur ($g'=1.92$), and $N_\textrm{e}$ is the number density of electrons. The $\gamma$-ray energy dependence of the analyzing power is plotted in Fig.~\ref{Edep_Pa}. The analyzing power for a 6.6-MeV $\gamma$ ray was estimated to be $P_a=1.89\%$. The change in the analyzing power between 2 and 7~MeV was also estimated to be approximately 6\%. 

\begin{figure}[htbp]
	\centering
		\includegraphics[clip,width=8cm]{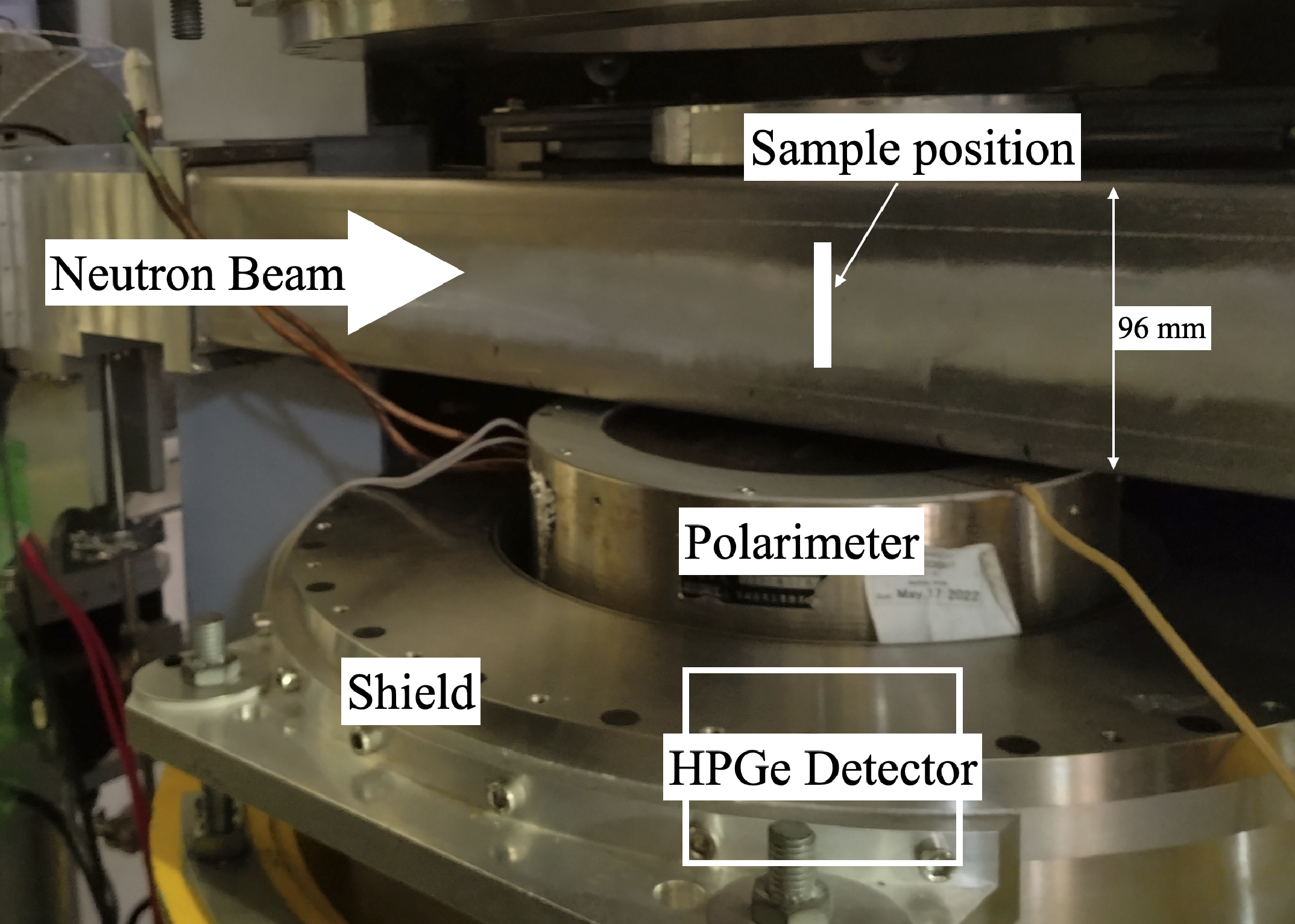}
	  \caption{Photograph of the polarimeter installed into ANNRI.}\label{installannri}
\end{figure}

Figure~\ref{installannri} shows a photograph taken after installing the polarimeter at ANNRI. The $\gamma$ rays transmitted through the polarimeter were detected using the HPGe detector. The HPGe detector has a hexagonal shape, and the size is approximately 70~mm in diameter and 78~mm in length. The details of the HPGe detector are given in Ref.~\cite{Takada}. The $\gamma$-ray detection efficiency was estimated by considering the transmission of the polarimeter using GEANT4 simulation~\cite{GEANT4}.
The photo peak efficiency was obtained to be $0.010\%$ for a 5-MeV $\gamma$ ray using the same method in Ref.~\cite{Takada}. This efficiency was approximately 0.2 times the detection efficiency without a polarimeter of 0.046\% due to the attenuation by the polarimeter. Installing the polarimeter at ANNRI made it possible to measure circular polarization with a full absorption peak efficiency of $0.010\%$ in a pulsed neutron source for the first time.

\section{Measurements of performance indicators for the polarimeter at UVSOR}

\subsection{Measurement procedure}
\begin{figure*}[htbp]
	\centering
		\includegraphics[clip,width=16cm]{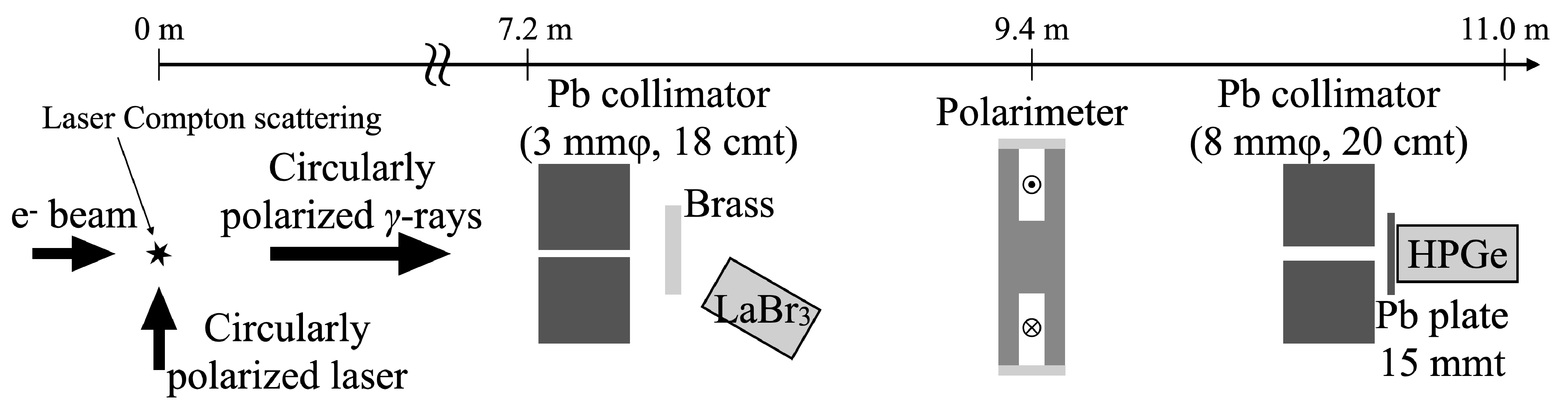}
	  \caption{Schematic view of the experimental setup at UVSOR facility. The circularly polarized $\gamma$-ray beam was produced using the laser Compton scattering (LCS) with circularly polarized laser light. The $\gamma$ rays transmitting the polarimeter were measured using the HPGe detector. The $\gamma$-ray flux incident on the polarimeter was monitored by measuring the $\gamma$ rays scattered by the brass plate using a LaBr$_3$ detector.}\label{UVSORsetup}
\end{figure*}

The transmission of circularly polarized $\gamma$ rays was measured at BL1U in the UVSOR~\cite{Katoh} facility to evaluate the analyzing power and magnetic hysteresis. Figure~\ref{UVSORsetup} shows the experimental setup.
The circularly polarized pulsed $\gamma$-ray beam was produced using laser Compton scattering (LCS). The maximum $\gamma$-ray energy was 6.6~MeV. The details of the generation of circularly polarized $\gamma$ rays are provided in Ref.~\cite{Taira}. As a $\gamma$-ray flux monitor, a LaBr$_3$ detector was used to measure the Compton scattered events with a 3-mm-thick brass plate between the collimator and polarimeter. The $\gamma$ rays penetrating the polarimeter were detected using an HPGe detector 90.3~mm in diameter and 83.5~mm in length. A lead (Pb) collimator was installed in front of the HPGe detector to reduce the small-angle scattering background in the polarimeter. An additional 15-mm-thick Pb plate was placed in front of the HPGe detector to reduce the pile-up events of the HPGe detector signals. 
The number of transmitted $\gamma$ rays for left- and right-handed circular polarizations was measured by changing the current applied to the polarimeter from $-6$~A to $6$~A. Furthermore, a laser-off measurement was performed to estimate the background from environmental radiation and the Bremsstrahlung of storage electrons.

\subsection{Analysis}
\begin{figure}[htbp]
	\centering
		\includegraphics[clip,width=8cm]{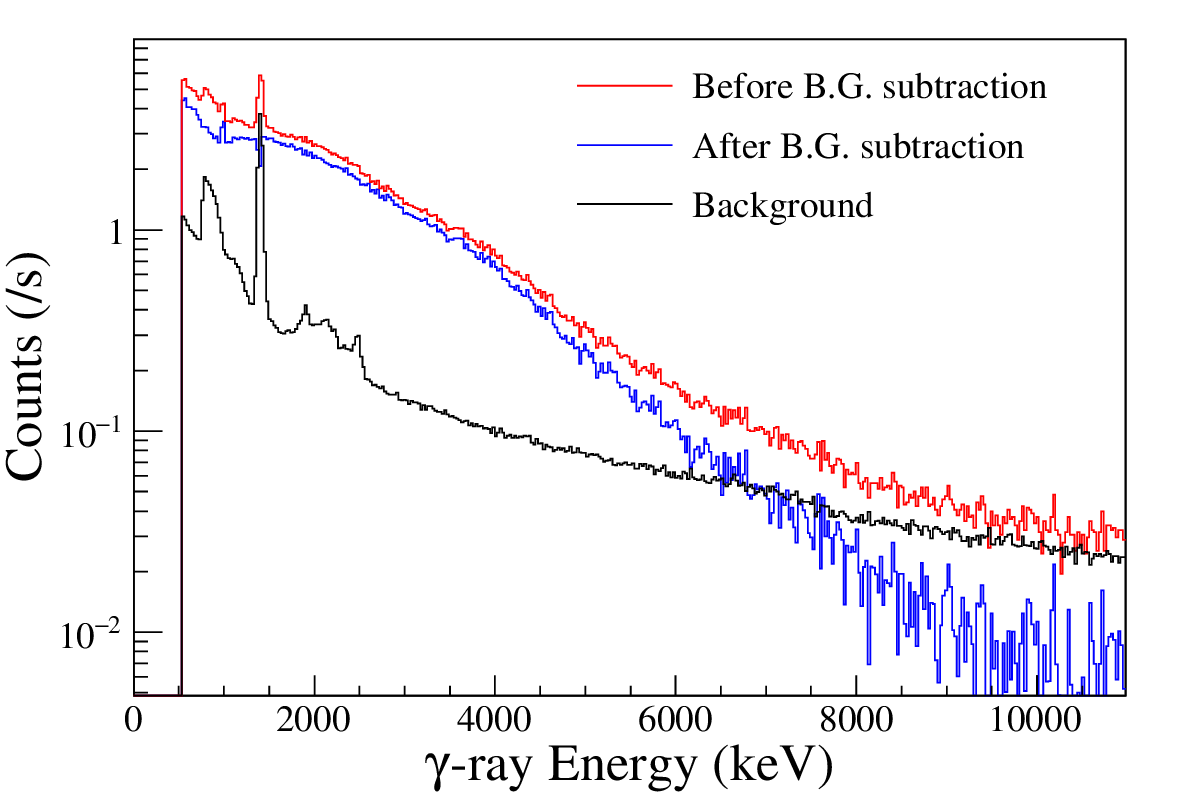}
	  \caption{Pulse height spectrum of the LaBr$_3$ detector. The flux was defined as the integral value from 2.0 to 6.7~MeV.}\label{fluxmonitorUVSOR}
\end{figure}

Figure~\ref{fluxmonitorUVSOR} shows the pulse height spectrum of the LaBr$_3$ detector. The background is the result of the laser-off measurement. The structures from 0.7 to 2.5~MeV results from the self-radioactivity of the LaBr$_3$ detector.
The $\gamma$-ray flux, $N_{\textrm{LaBr}_3}$, was determined using an integral value from 2.0 to 6.7~MeV. The transmitted events were normalized to the incident $\gamma$-ray intensity measured at each current applied to the polarimeter.

\begin{figure}[htbp]
	\centering
		\includegraphics[clip,width=8cm]{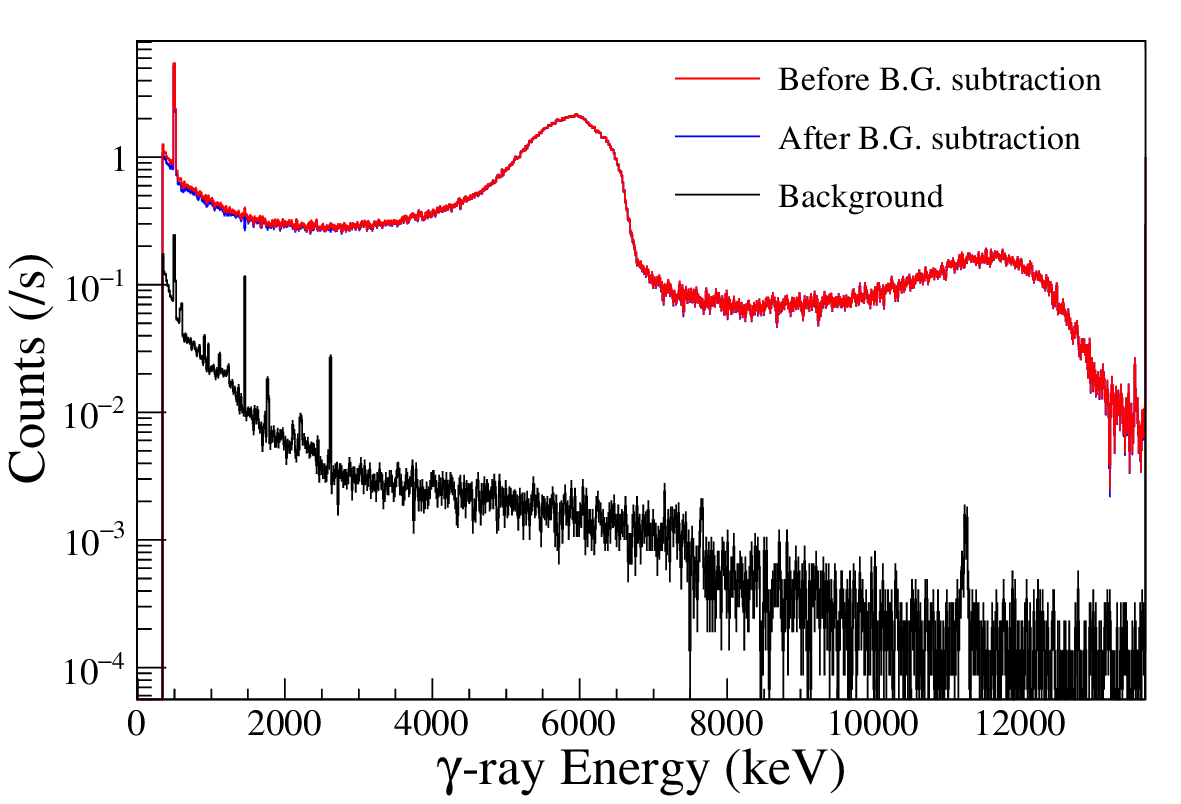}
	  \caption{Gamma-ray spectrum of the Ge detector with the background. The spectra before and after background subtraction almost overlap because of the low background.}\label{gamspeUVSOR}
\end{figure}

Figure \ref{gamspeUVSOR} shows the $\gamma$-ray energy spectrum of the HPGe detector with the background of laser-off measurement.
The peak around 12 MeV was caused by pile up. The origin of 11.2-MeV peak in the background could not be specified, but it does not affect the results. The total counts of the HPGe detector corrected for pile-up events were obtained using the following procedure. According to Ref.~\cite{Kondo}, the number of pile-up $\gamma$-rays follows a Poisson distribution:
\begin{equation}
\label{poas}
    P_\lambda(m)=\frac{\lambda^m e^{-\lambda}}{m!},
\end{equation}
where $\lambda$ relates to the average number of $\gamma$ rays incident on the HPGe detector per pulse, and $m$ is the number of pile-ups. The number of single and double $\gamma$-ray detection events is written as
\begin{equation}
\label{singlehit}
    N_1=P_\lambda(1)\varepsilon+P_\lambda(2)2\varepsilon(1-\varepsilon),
\end{equation}
and
\begin{equation}
\label{doublehit}
    N_2=P_{\lambda}(2)\varepsilon^2,
\end{equation}
where $N_1$ is the count of the single $\gamma$-ray events, which is the integral value between 200 and 6,620~keV; $N_2$ is the count of the double $\gamma$-ray events, which is an integral value between 7,120 and 13,240~keV. $\varepsilon$ is the detection efficiency. The detection efficiency was estimated to be 80\% using an EGS5 simulation. From Eqs.~(\ref{poas})-(\ref{doublehit}), the parameter, $\lambda$, is obtained as follows:
\begin{equation}
    \lambda=\frac{2N_2}{\varepsilon N_1-2N_2(1-\varepsilon)}.
\end{equation}
 The total count, considering up to the fourth hit, can be calculated as
 \begin{equation}
     N_\textrm{HPGe}=\sum_{i=1}^{4}P_\lambda(i)=P_\lambda(1)\left(1+\frac{\lambda}{2}+\frac{\lambda^2}{6}+\frac{\lambda^3}{24}\right)
 \end{equation}
 where $P_\lambda(1)$ is calculated as
 \begin{equation}
     P_\lambda(1)=\frac{N_1}{\varepsilon}-\frac{N_2}{\varepsilon^2}2(1-\varepsilon)
 \end{equation}
 from Eqs.~(\ref{singlehit}) and (\ref{doublehit}). The corrected count proportional to the transmission of the polarimeter, $S$, was defined as
 \begin{equation}
 \label{eq:transcount}
     S=\frac{N_\textrm{HPGe}}{N_{\textrm{LaBr}_3}}.
 \end{equation}

\subsection{Results}

\begin{figure}[htbp]
	\centering
		\includegraphics[clip,width=8cm]{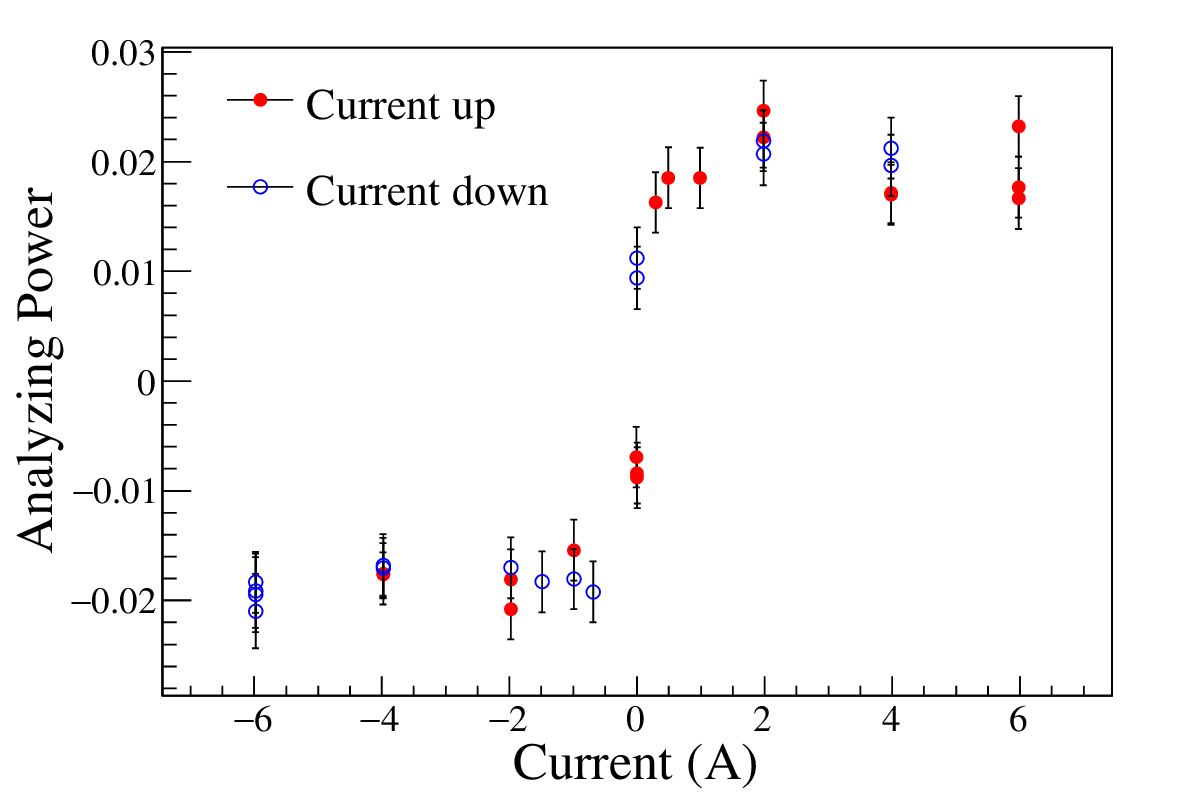}
	  \caption{Analyzing power for each current applied to the polarimeter. The closed and opened points indicate whether the measurement is for increasing or decreasing the current, respectively. The uncertainty was deduced from statistical errors.}\label{analyzingpower}
\end{figure}

The corrected counts, $S_\textrm{r}$ and $S_\textrm{l}$, for the right- and left-handed circularly polarized $\gamma$-ray beam were obtained from Eq.~(\ref{eq:transcount}), and the analyzing power was calculated as
\begin{equation}
P_a=\frac{S_\textrm{r}-S_\textrm{l}}{S_\textrm{r}+S_\textrm{l}},
\end{equation}
based on Eq.~(\ref{eq:Padef1}).
Figure~\ref{analyzingpower} shows the analyzing power determined for each current. The closed and opened points indicate whether the measurement was for increasing or decreasing the current of the polarimeter, respectively. The uncertainty was deduced from propagating statistical errors.
A difference at 0~A observed in the analyzing power is due to the effect of magnetic hysteresis.
 
The analyzing power, saturated above $2$~A and below $-2$~A, was obtained to be $2.02\%\pm0.08\%$ and $-1.83\%\pm0.08\%$, respectively. The analyzing powers for the positive and negative currents differ by approximately 5\% because the internal magnetic field differs slightly, particularly in the return yoke. This effect was considered the systematic uncertainty of the analyzing power, 0.10\%. Consequently, the averaged analyzing power was determined as $1.92\%\pm0.06\%\pm0.10\%$, consistent with the estimated value of $1.89\%$ described in Sec. 2. 

\section{Measurements of the circular polarization at ANNRI}

\subsection{Circular polarization of $\gamma$ rays from $^{32}$S($\vec{\textrm{n}}$,$\gamma$)$^{33}$S reactions}
As a demonstration of circular polarization measurements with a pulsed neutron source, such as the one in MLF, the $\gamma$-ray circular polarization was measured in $^{32}$S($\vec{\textrm{n}}$,$\gamma$)$^{33}$S reactions. The circular polarization of $\gamma$ rays emitted after a polarized neutron capture reaction is described as \cite{Biedenharn}
\begin{equation}
\label{Eq:Pg}
P_\gamma=P_\textrm{n}R\cos\Theta,
\end{equation}
where $P_\textrm{n}$ is the neutron polarization ratio, $\Theta$ is the angle between the directions of neutron polarization and the emitted $\gamma$-ray momentum, and $R$ is the coefficient calculated from the nuclear spin of the compound state, of the initial state, and of the final state. Since the $R$-value for the transition emitting 5.4-MeV $\gamma$ rays in the $^{32}$S($\vec{\textrm{n}}$,$\gamma$)$^{33}$S reactions is $R=-1/2$~\cite{Konijnenberg}, this reaction was used to determine the analyzing power in previous experiments~\cite{MikePol}. 

The $\gamma$-ray circular polarization is experimentally obtained as
\begin{equation}
\label{Eq:Pa}
P_\gamma^\textrm{exp}=\frac{1}{P_a}\frac{N^\textrm{up}_\textrm{r}-N^\textrm{down}_\textrm{r}}{N^\textrm{up}_\textrm{r}+N^\textrm{down}_\textrm{r}}
\end{equation}
where $P_a$ is the analyzing power and $N^\textrm{up}_\textrm{r}$ and $N^\textrm{down}_\textrm{r}$ are the detected counts for up- and down-polarized neutrons, respectively. The analyzing power has already been determined in the UVSOR experiment, but here it is necessary to consider the effects of the solid angle. Thus, we reevaluated the analyzing power using the 5.4-MeV $\gamma$ ray. Moreover, the circular polarization of other $\gamma$ rays was obtained and compared with the theoretical calculation.

\subsection{Measurement procedure}

\begin{figure*}[htbp]
	\centering
		\includegraphics[clip,width=16cm]{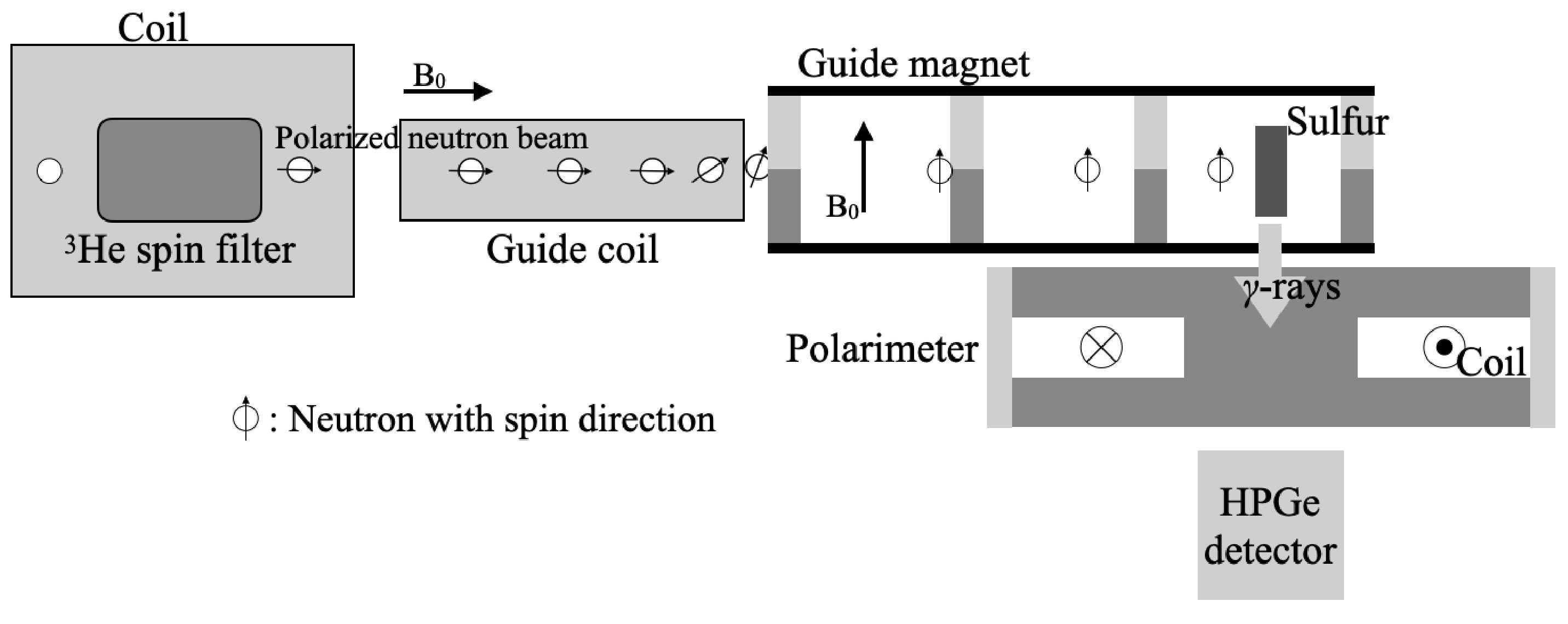}
	  \caption{Schematic view of the experimental setup in ANNRI. The longitudinally polarized neutrons were produced using a $^{3}$He spin filter, and a guide magnet turned the polarization direction up (down). The current of the polarimeter was fixed at 3.98 A. Li-glass neutron detectors were installed at 7-m downstream of the sulfur sample to monitor the neutron polarization. The angle, $\Theta$, in Eq.~(\ref{Eq:Pg}) equals $180^\circ$ for up-polarized neutrons and $0^\circ$ for down-polarized neutrons.}\label{setup}
\end{figure*}

Figure~\ref{setup} shows the experimental setup.
The HPGe detector was at a flight length of 21.5~m. The $\gamma$-ray polarimeter was installed between the HPGe detector and the sulfur sample. In this measurement, the angle, $\Theta$, in Eq.~(\ref{Eq:Pg}) equals $180^\circ$ for up-polarized neutrons and $0^\circ$ for down-polarized neutrons. Longitudinally polarized neutrons were produced using a $^{3}$He spin filter~\cite{Okudaira_He}, and a guide magnet turned the polarization direction up (down). Li-glass detectors were installed at a flight length of 28.2 m, and transmitted neutrons were measured to correct the neutron polarization and neutron beam intensity. There are two types of Li-glass detectors: $^{6}$Li-enriched $^{6}$Li-glass and $^{7}$Li-enriched $^{7}$Li-glass. $^{6}$Li-glass has a high sensitivity to neutrons, whereas $^{7}$Li-glass has a slight sensitivity to neutrons. The $\gamma$-ray background can be removed by subtracting the spectrum of $^{7}$Li-glass from that of $^{6}$Li-glass. Natural sulfur powder was pressed and shaped into a $\phi$24 mm and 1.3-cm-thick tablet. Total of 28.9 and 29.1 hours were measured for the up- and down-polarization directions, respectively.

\begin{figure}[htbp]
	\centering
		\includegraphics[clip,width=8cm]{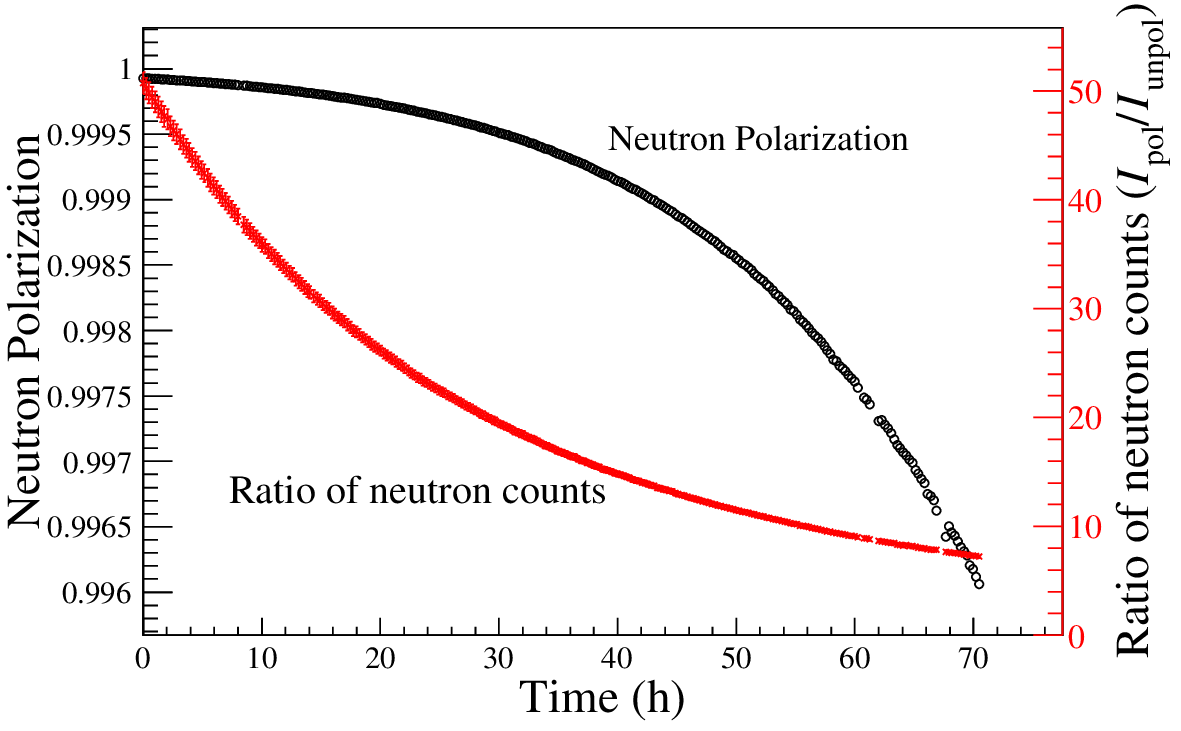}
	  \caption{Time dependence of the neutron polarization ratio and the ratio of the transmitted neutron counts averaged from 3 meV to 25 meV. The ratio of neutron counts corresponds to the number of neutrons at the sample position.}\label{polratio}
\end{figure}

Figure~\ref{polratio} shows the time dependence of the neutron polarization ratio and the ratio of the transmitted neutron counts of the $^{3}$He spin filter.
The polarization ratio was obtained from the ratio of the transmitted neutron counts of the polarized cell to that of the unpolarized cell. The details of the polarization determination are given in Refs.~\cite{Yamamoto,Endo2022}. The ratio of transmitted neutron counts was obtained by dividing the number of events from 3 to 25 meV in the polarized cell by that in the unpolarized cell. Neutron counts at the sample position decrease with time because of the depolarization of $^{3}$He nuclei.

\subsection{Analysis and results}

\begin{figure}[htbp]
	\centering
		\includegraphics[clip,width=8cm]{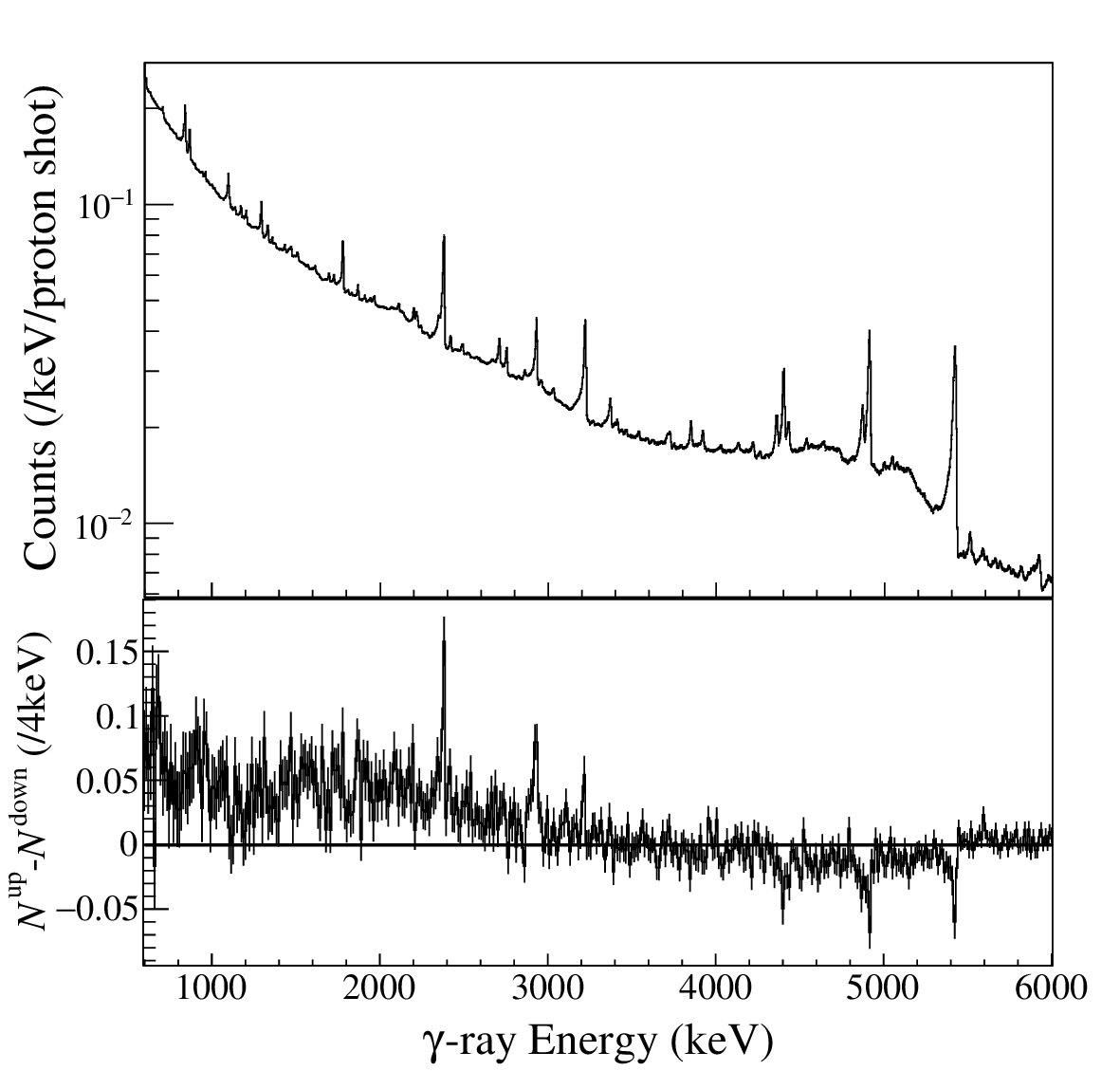}
	  \caption{Gamma-ray energy spectra for the up-polarization gated neutron energy from 3 to 25~meV (top), and the difference ($N^\textrm{up}-N^\textrm{down})$ between up- and down-polarization (bottom). A noticeable difference can be found in the 5.4-MeV and some other $\gamma$ rays.}\label{gamspe}
\end{figure}

\begin{figure}[htbp]
	\centering
		\includegraphics[clip,width=8cm]{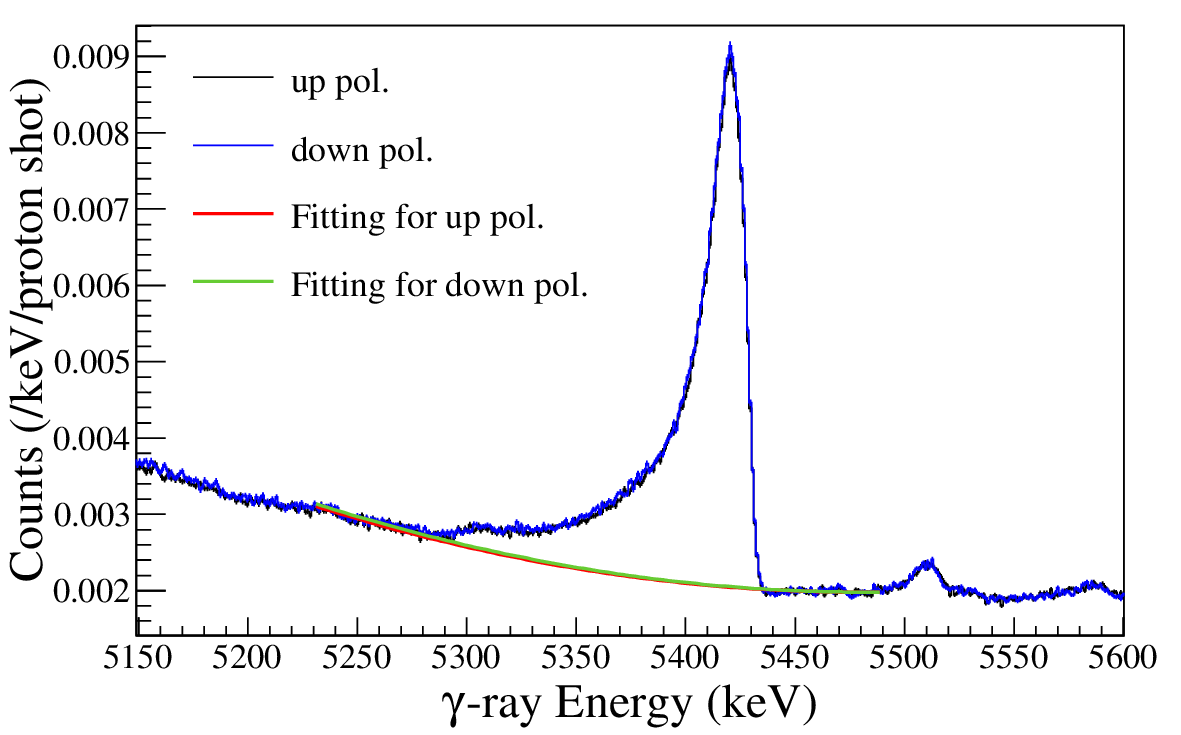}
	  \caption{Gamma-ray energy spectra around the 5.4-MeV full absorption peak. To obtain the net counts of the peak, the backgrounds were estimated using a 3-rd polynomial function fit shown as the pink and green lines.}\label{gamFit}
\end{figure}

Figure \ref{gamspe} shows the $\gamma$-ray energy spectrum for the up-polarization measurements of gated neutron energy from 3 to 25 meV and the difference between the up- and down-polarization measurements. Besides the 5.4-MeV $\gamma$ rays, noticeable differences were observed in some $\gamma$ rays, and these are discussed in Sec.~5.
The backgrounds were estimating by a 3-rd polynomial function fit, and the net counts of 5.4-MeV $\gamma$ rays were deduced for the full absorption, single escape, and double escape peaks. Figure~\ref{gamFit} shows the background fit for the full absorption peak. The net count, $N_\textrm{r}$, was obtained by subtracting the integral of the fitting result from that of the $\gamma$-ray spectrum. The ranges of the integral regions were defined as the full width at quarter maximum (FWQM).

The asymmetry was defined as
\begin{equation}
    A=\frac{N^\textrm{up}_\textrm{r}-N^\textrm{down}_\textrm{r}}{N^\textrm{up}_\textrm{r}+N^\textrm{down}_\textrm{r}}.
\end{equation}
To correct the decreasing beam intensity due to the depolarization of the $^{3}$He nuclei and the neutron polarization, the corrected asymmetry $A'$ was defined as
\begin{equation}
    A'=\frac{(T^\textrm{up}+T^\textrm{down})A-(T^\textrm{up}-T^\textrm{down})}{(P^\textrm{up}_\textrm{n}T^\textrm{up}_\textrm{n}+P^\textrm{down}_\textrm{n}T^\textrm{down}_\textrm{n})-A(P^\textrm{up}_\textrm{n}T^\textrm{up}_\textrm{n}-P^\textrm{down}_\textrm{n}T^\textrm{down}_\textrm{n})},
\end{equation}
where $P_\textrm{n}^\textrm{up}$ and $P_\textrm{n}^\textrm{down}$ are the neutron polarization. $T^\textrm{up}_\textrm{n}$ and $T^\textrm{down}_\textrm{n}$ are the neutron transmissions monitored at the Li-glass detectors for each polarization direction. The neutron transmission corresponds to the neutron beam intensity. Table \ref{Pa_ANNRI} lists the corrected asymmetries for the 5.4-MeV $\gamma$ ray, including the single and double escape peaks. The asymmetry combined with these values was obtained as $A'=-(8.62\pm1.07)\times10^{-3}$. The analyzing power was calculated as $P_a=A'/R$ from Eqs.~(\ref{Eq:Pg}) and (\ref{Eq:Pa}), and $P_a=1.72\%\pm0.21\%$ was obtained for the 5.4-MeV $\gamma$ rays. Adding the systematic uncertainty due to the difference in the positive and negative currents obtained from the UVSOR experiments, the analyzing power was determined to be $P_a=1.72\pm0.23\%$. 

This result is slightly smaller than the expected value, $2.03\%$, from the UVSOR experiments, considering the energy dependence of the analyzing power in Fig.~\ref{Edep_Pa}. In the ANNRI experiments, since all $\gamma$ ray was not incident perpendicular to the polarimeter due to the solid angle of the detector from the sample, the analyzing power was reduced. Therefore, it is necessary to use the results of this sulfur measurement to calibrate the circular polarization measured in ANNRI.

\begin{table}[htbp]
\centering
\caption{Obtained asymmetries for the 5.4-MeV $\gamma$ rays.}\label{Pa_ANNRI}
\begin{tabular}{c|c}\hline
 Peak & Asymmetry $A'$ ($\times10^{-2}$)  \\ \hline
Full absorption & $-0.814\pm0.137$ \\
Single escape & $-0.824\pm0.200$ \\
Double escape & $-1.280\pm0.343$ \\ \hline
Averaged value & $-0.862\pm0.107$ \\\hline

\end{tabular}
\end{table}




\section{Discussion of the circular polarization of $\gamma$ rays in $^{32}$S($\vec{\textrm{n}}$,$\gamma$)$^{33}$S reactions}
In this section, we discuss the circular polarization of other $\gamma$-rays in $^{32}$S($\vec{\textrm{n}}$,$\gamma$)$^{33}$S reactions.

\subsection{Theoretical calculation}
When a neutron with an angular momentum of 0 is captured (s-wave resonance), the coefficient $R$ in Eq.~(\ref{Eq:Pg}) can be calculated~\cite{KOPECKY} by
\begin{eqnarray}
\label{R}
    R=&&\frac{1}{2}F_1\left(\frac{1}{2}\frac{1}{2}J_\textrm{i}J_\textrm{c}\right)\nonumber\\
    &&\times\frac{F_1(L_1L_1J_\textrm{f}J_\textrm{c})-2\delta F_1(L_1L_2J_\textrm{f}J_\textrm{c}) +\delta^2 F_1(L_2L_2J_\textrm{f}J_\textrm{c})}{1+\delta^2}, 
\end{eqnarray}
where $F_1$ is the coefficient for angular correlations tabulated in Refs.~\cite{Lawrence,Alder}. $J_\textrm{i}$, $J_\textrm{c}$, and $J_\textrm{f}$ are the spins of the initial, compound, and final states, respectively. $L_1$ and $L_2$ are the multipolarity of possibly mixed $\gamma$ rays, and $\delta$ is the mixing ratio. In the following, only $L_1=1$ and $L_2=2$ are considered. Furthermore, if the $\gamma$-ray transition is via an intermediate state (spin $J_\textrm{m}$) by unobserved $L_\textrm{m}$ radiation, Eq.~(\ref{R}) is written ~\cite{Lawrence} as
\begin{eqnarray}
\label{Eq:R2}
    R=&&\frac{1}{2}F_1\left(\frac{1}{2}\frac{1}{2}J_\textrm{i}J_\textrm{c}\right)U(m)\nonumber\\
    &&\times\frac{F_1(11J_\textrm{f}J_\textrm{m})-2\delta F_1(12J_\textrm{f}J_\textrm{m}) +\delta^2 F_1(22J_\textrm{f}J_\textrm{m})}{1+\delta^2}, 
\end{eqnarray}
where 
\begin{eqnarray}
    U(m)=&&(-1)^{J_\textrm{m}-J_\textrm{c}-L_\textrm{m}}\sqrt{(2J_\textrm{m}+1)(2J_\textrm{c}+1)}\nonumber\\
    &&\times W(J_\textrm{c}J_\textrm{c}J_\textrm{m}J_\textrm{m};1L_\textrm{m}),
\end{eqnarray}
where $W$ is the Racah coefficient, calculated using the Wigner 6-j symbol as
\begin{equation}
   W(abcd;ef)=(-1)^{a+b+c+d}\left\{
\begin{matrix}
a & b & e \\
d & c & f \\
\end{matrix}
\right\}.
\end{equation}
Here, the mixing of multipolarity in the intervening transition is not considered. 

\begin{figure}[htbp]
	\centering
		\includegraphics[clip,width=8cm]{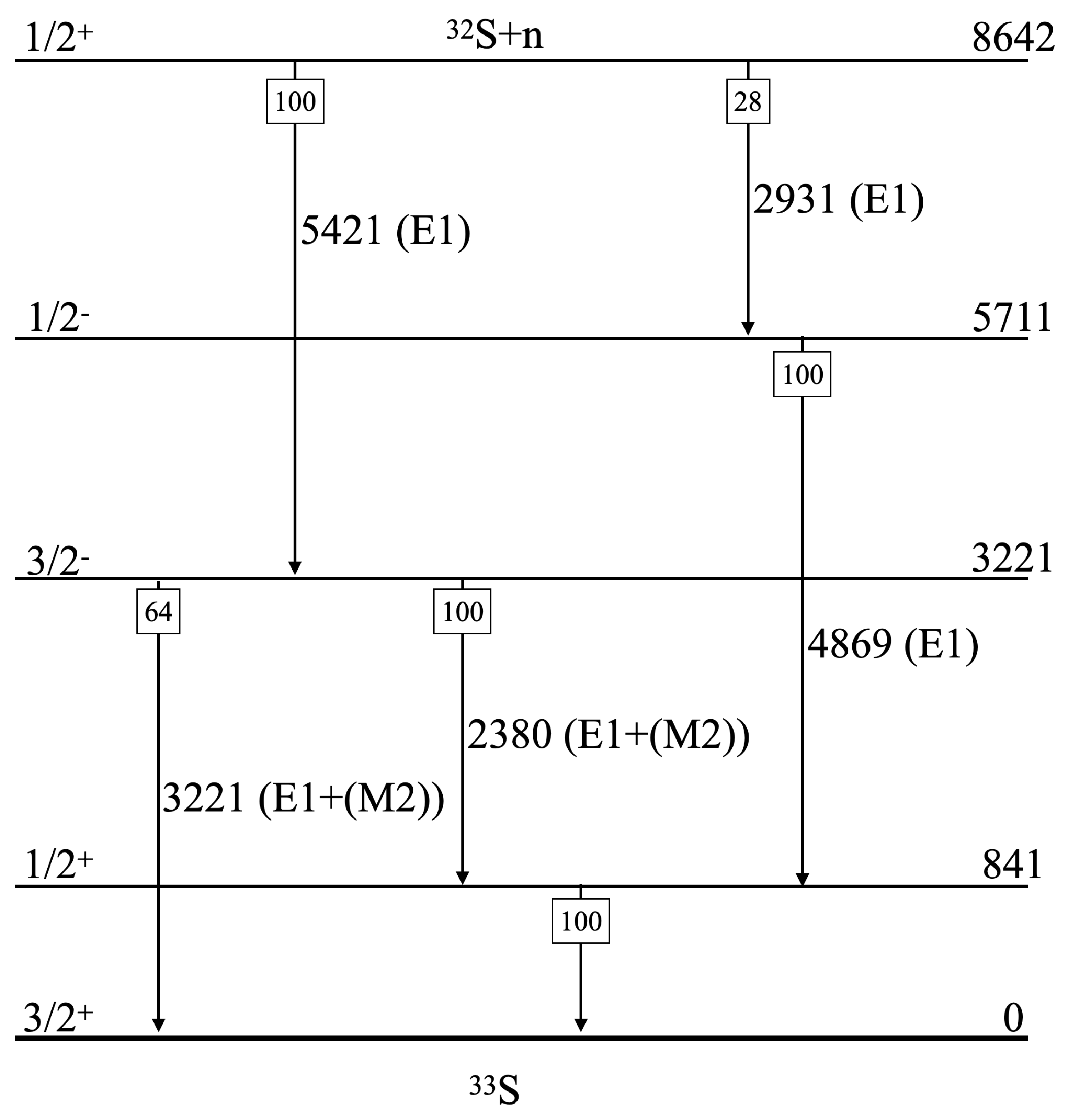}
	  \caption{Decay scheme showing only the major transitions observed in the $^{32}$S(n,$\gamma$)$^{33}$S reaction taken from Ref.~\cite{ENSDF_S}. The excitation energies and the $\gamma$-ray transition energies are shown in the unite of keV. The number in the square on the arrows indicates the relative transition strength with the strongest transition being 100.} \label{S_Decay}
\end{figure}

Figure~\ref{S_Decay} displays the decay scheme, showing only the major transitions from the $^{32}$S$+$n compound state. Table~\ref{R_Hikaku} lists the calculated $R$-values with $\delta=0$ in each transition using Eq.~(\ref{Eq:R2}). For example, the $R$-value of a 3.2-MeV $\gamma$ ray is deduced here. The spins of the initial, compound, intermediate, and final states are $J_\textrm{i}=0$, $J_\textrm{c}=1/2$, $J_\textrm{m}=3/2$, and $J_\textrm{f}=1/2$, respectively, and the 5.4-MeV and 3.2-MeV transitions are considered as E1 transitions. Since $F_1\left(\frac{1}{2}\frac{1}{2}0\frac{1}{2}\right)=-2$, $U(m)=0.745$, and $F_1\left(11\frac{1}{2}\frac{1}{3}\right)=-0.447$, the $R$-value becomes $0.333$.

\begin{table}[htbp]
\centering
\caption{The $R$-values for each transition obtained from the theoretical calculations and the experiments. The calculation only considers the multipolarity$=1$ transition. The first and second uncertainties in the experimental value are derived from the uncertainties of the statistics and the analyzing power, respectively.}\label{R_Hikaku}
\begin{tabular}{c|cc}\hline

 $\gamma$-ray Energy & $R$ from calculation & $R$ from experiment \\ \hline\

 5420 keV & -0.5 & $-0.5$ (Using Normalization)\\
 3221 keV & 0.33 & $0.52\pm0.13\pm0.07$ \\
 2380 keV & 0.83  & $0.71\pm0.10\pm0.10$ \\
 2930 keV & 1.00  & $1.05\pm0.21\pm0.14$ \\
 4869 keV & -0.33 & $-0.78\pm0.28\pm0.11$ \\ \hline

\end{tabular}
\end{table}

\subsection{Comparison with the experimental results}
The same analysis for the 5.4-MeV $\gamma$ ray in Sec.~4 was applied to the other $\gamma$ rays, and the $R$-values for each transition were derived and are listed in Table~\ref{R_Hikaku}. The integral regions used to derive the net count of the photo peak were defined as FWQM, the same as the 5.4-MeV analysis, however, the full width at half maximum was applied to the 4.9-MeV $\gamma$ ray because the peak count was low.  In the derivation of the $R$ values, the analyzing power was calculated using the result for 5.4-MeV $\gamma$ rays and the energy dependence shown in Fig.~\ref{Edep_Pa}. The experimental results correlate with the theoretical calculations within 1.5~$\sigma$ for all transitions.

\begin{figure}[htbp]
  \begin{minipage}{0.5\hsize}
  \begin{center}
  \subfloat[\label{RD_2380}For $2380$-keV $\gamma$-ray]{\includegraphics[width=8cm]{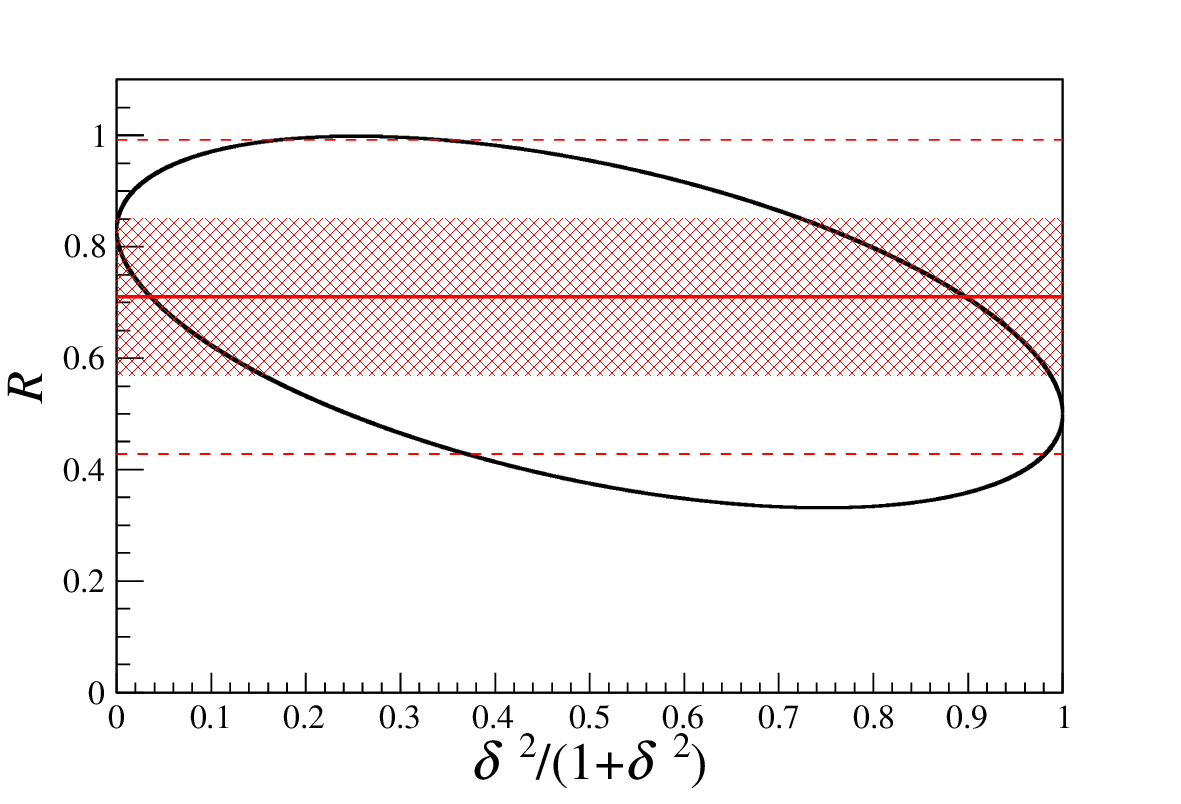}}
  \end{center}
 \end{minipage}
 \begin{minipage}{0.5\hsize}
  \begin{center}
    \subfloat[For $3221$-keV $\gamma$-ray]{\includegraphics[width=8cm]{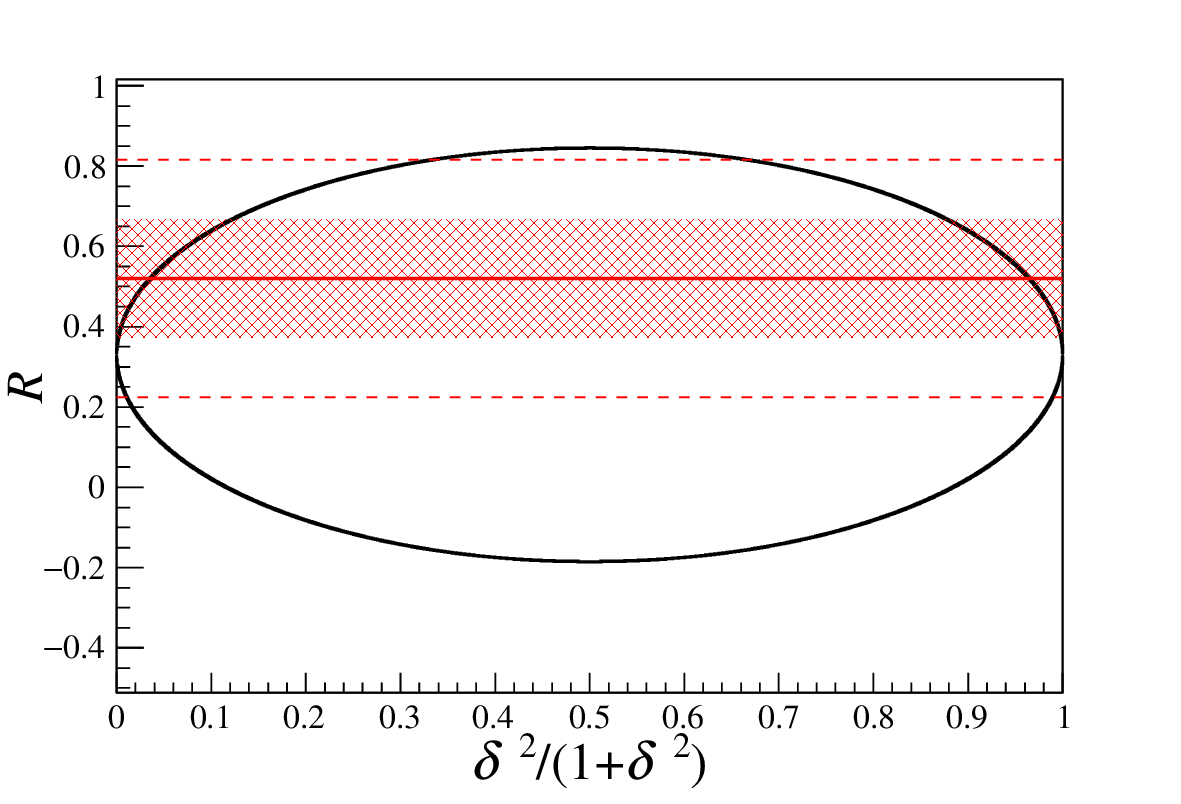}}
  \end{center}
 \end{minipage}
 \caption{\label{RD_each} The $R$-value for each $\gamma$ ray as a function of $\delta^2/(1+\delta^2)$. The colored region represents the $R\pm1\ \sigma$ region, and the dotted lines represent the $R\pm2\ \sigma$ values.}
 \end{figure}

It is also possible to derive the mixing ratio of the transition of multiplicity$=2$ (i.e., in this case, the mixing ratio of the M2 transition in the E1 transition) from this result. In the following, the mixing ratio, $\delta$, was derived for the 2.4- and 3.2-MeV $\gamma$ rays as a demonstration. The $R$-values considering the $L=2$ transition are $R=\frac{0.50\delta^2-0.58\delta+0.83}{1+\delta^2}$ and $R=\frac{0.33\delta^2-1.03\delta+0.33}{1+\delta^2}$ for the 2.4- and 3.2-MeV $\gamma$ rays from Eq.~(\ref{Eq:R2}). Figure \ref{RD_each} displays the $R$-value for the 2.4- and 3.2-MeV $\gamma$ rays as a function of $\delta^2/(1+\delta^2)$. The colored region represents the $1\ \sigma$ region, and the dotted lines represent the $\pm2\ \sigma$. The $\delta=0.43^{+0.31}_{-0.43}$ and $\delta=0.19^{+0.18}_{-0.15}$ for the 2.4- and 3.2-MeV transitions were obtained. Two solutions were produced; however, mixing M2 transitions is generally so small that the one with the smaller $\delta$ is described. Although significant mixing of the M2 transition was not observed because of the large uncertainty, the mixing ratio can be determined by measuring the circular polarization in ANNRI.

It is expected to determine the spin, $J$, and the mixing of transitions by performing similar experiments on other nuclei and improving the accuracy in the future. We plan to apply this method to determine the spin of the final state of $^{139}$La$+$n, which is crucial for verifying the s-p mixed model.

\section{Conclusions}
Measurements of the circular polarization of $\gamma$ rays emitted from neutron capture reactions provide valuable information for studying nuclear physics. The $\gamma$-ray polarimeter was developed as a $\gamma$-ray circular polarization measurement system installed in a highly intense pulsed neutron source. The analyzing power and magnetic hysteresis, used as performance indicators, were measured using a circularly polarized $\gamma$-ray beam at BL1U in the UVSOR facility. The analyzing power was obtained to be $1.92\%\pm0.06\%\pm0.10\%$ for the $6.6$-MeV $\gamma$ rays. It was confirmed that the current applied to the polarimeter was sufficiently saturated when it exceeded $2$~A.

 Furthermore, as a demonstration of circular polarization measurement in the pulsed neutron source, the circular polarization of $\gamma$ rays emitted from $^{32}$S($\vec{\textrm{n}}$,$\gamma$)$^{33}$S reactions with polarized neutrons was measured at ANNRI of the MLF in J-PARC. The analyzing power was found to be $1.72\%\pm0.23\%$ from the $5.4$-MeV $\gamma$ rays. The circular polarization of other $\gamma$ rays normalized to this value was consistent with the theoretical calculations. It has been possible to measure the circular polarization of $\gamma$ rays from neutron capture reactions in a pulsed neutron source for the first time. Various experiments in nuclear physics using circular polarization of neutron capture reactions are expected to be performed at ANNRI in the future.

\begin{acknowledgements}
The authors would like to thank the staff of beamline 04, ANNRI, for the maintenance of the germanium detectors, and MLF and J-PARC for operating the accelerators and the neutron production target. The authors would also like to thank the staff of beamline 10, NOBORU, where the exploratory experiment was performed. The neutron experiments at the MLF of J-PARC were performed under the user program (Proposals No. 2021B0385, 2022A0230, 2022B0187, 2023A0129). The $\gamma$-ray experiments were performed at the BL1U of the UVSOR Synchrotron Facility with the approval of the Institute for Molecular Science (IMS), NINS (Proposals No.22-602 and 22-803). This work was supported by the Neutron Science Division of KEK as an S-type research project with program number 2018S12. This work was partially supported by JSPS KAKENHI Grant Nos. JP19K21047, JP20K04007, JP20K14495, and JP21K04950.

\end{acknowledgements}

\bibliographystyle{spphys}       
\bibliography{cas-refs}   

\end{document}